\documentclass{sig-alternate}
\usepackage{url}
\begin{document}
\title{It Takes a Socio-Technical Ecosystem}
\numberofauthors{2}

\author{
\alignauthor
John D. McGregor\\
       \affaddr{Strategic Software Engineering Research Group}\\
       \affaddr{Clemson University}\\
       \affaddr{Clemson, SC 29634}\\
       \email{johnmc@clemson.edu}
\alignauthor
J. Yates Monteith\\
       \affaddr{Strategic Software Engineering Research Group}\\
       \affaddr{Clemson University}\\
       \affaddr{Clemson, SC 29634}\\
       \email{jymonte@clemson.edu}
}
\maketitle

\begin{abstract}
There are both technical and social issues regarding the design of 
sustainable scientific software. Scientists want continuously evolving
systems that capture the most recent knowledge while developers and architects
want sufficiently stable requirements to ensure correctness and efficiency. A
socio-technical ecosystem provides the environment in which these issues
can be traded off.  

\end{abstract}

\section{Introduction}\label{intro}
Software is increasingly the tool of choice for tasks in scientific research,
ranging from simulations replacing expensive, high risk or dangerous experiments to
collecting, transforming and analyzing vast amounts of data. Scientific software has reached
levels of complexity rivaling those of modern skyscrapers, yet the software
is often constructed by people with little or no background in the design or
sustainment of software products. The term ``software ecosystem''~\cite{bosch3} has been
used to describe the interactions among organizations that collaborate on the
development of a common platform and compete through the products built from
that platform. The term hides an essential element in the development of
scientific software - the people who provide scientific and software expertise.
``Socio-technical ecosystems''~\cite{Geels2004897} clearly express the idea that the people part
of software development and sustainment is of equal importance to the software
part of development and sustainment. 

The lead author began creating physics simulations as an undergraduate and
went on to spend some time at  a national laboratory. Most recently a National
Science Foundation grant allowed both authors to focus on software development
in scientific research. We have examined research groups ranging from the
students of an individual researcher to large multi-institutional groups as well as an
international network of researchers in an effort to provide recommendations on 
scientific research software ecosystems~\cite{Monteith2014}. In this very short position paper we will
summarize several observations resulting from our investigations of ecosystems.

\section{Observations}\label{observations}
{\bf Recognize the role of system architecture}
Create a system architecture, even a simple one, can provide both structure to both the
software and the teams implementing the system. On the other hand, being unaware
or ignorant of system architecture can result in software that is hard to maintain, 
change and extend~\cite{ferenbaugh2013experiments}.  The architecture guides the
identification of critical sections that should be assigned to the most
experienced people. Following the architecture also optimizes the lines of
communication among the teams.  Recognizing and being aware of the role of 
architecture is the difference between a software system that extends, scales and is flexible
over time~\cite{Amorim1, amorim2, amorim3} and an software tool that remains in production 
past its expiration date.

{\bf Structure the community and establish governance}
Structure the community to be compatible with the system architecture.
The means of structuring should encourage interactions between representatives of 
different organizations~\cite{ferenbaugh2013experiments}.  Establishing a governance 
system that recognizes the contributions of the software engineers as well as the 
scientists is key to sustainable scientific software. The ecosystem should be structured to 
prevent the dominance from a single large contributer.

{\bf Distinguish between stable and rapidly evolving knowledge} 
Separate stable knowledge from the new knowledge being created through
research results. The system architecture should be modular in places where it is  likely
to change and that is optimized in those areas where change is less
likely. Using a platform architecture, the stable knowledge is captured in the
platform and new knowledge is captured in the extensions that complete a product.
In the NAMD molecular dynamics suite the maintainers 
keep core functionality stable while providing extension mechanisms through scripting
engines that allow users to develop custom and rapidly evolving routines and analyses~\cite{schulten}.

{\bf Make carpentry as important as results}
Make code quality as critical as research results. The science team focuses on a moving frontier. they conduct experiments, record
results, publish findings, and push forward. Under the pressure of paper
deadlines and funding reviews, code quality will have fallen by the wayside, reflected
in a lack of unit tests, documentation, commenting or technical debt.  Key to ensuring the 
long lived success and fitness of the software being developed, the software team 
needs to schedule time to pay back that debt by doing some cleanup work. There is 
value in engaging in software engineering practices such as regression testing, 
unit testing and test-driven development~\cite{clune2014testing}, as well as code 
review~\cite{petre2014code}. Making test artifacts, architecture, source code and 
requirements correct with respect to the final iteration is a useful investment to 
the longterm sustainability and success of the project. 

{\bf Designate gatekeepers to maintain integrity of the code base}
Protect the integrity of the core platform by being judicious in the inclusion of new 
contributions.  The Eclipse model as well as that of many projects gives special status to a few
people who spend some of their effort reviewing code submitted by others and who
organize the contributions in satisfactory ways.  This is necessary when a large
percentage of the developers are graduate students, particularly when the students 
are not software engineering students~\cite{schulten, Monteith2014}. The meritocracy 
governance approach of Eclipse is carried into the fact that gatekeepers are voted into the
position based on the quality of their contributions.

{\bf Maintain development roadmaps}
Build a roadmap to look into the future.  Building a roadmap is a matter of 
thinking ahead. Scientists are thinking
about sequences of experiments and software should likewise be planned ahead to
guide the architects and provide a clear path of development moving towards the future.
Several consortia are aware of the benefits of utilizing a roadmap to provide clear 
planning on the developmental trajectory on software, including
the Apache Foundation and the Eclipse Foundation.  Within the Eclipse Science Working
Group (ESWG), the Chemclipse and DAWNSci projects follow 
the Eclipse Development Process (EDP) roadmap as an incubating project in order to 
be integrated with the Eclipse Technology Project.  Similar to the EDP, the Apache Software Foundation
provides a process roadmap known as The Apache Way. These processes provides a roadmap from 
the inception and proposal of a new project through maturity to integration and archiving in their respective
ecosystems.

{\bf Develop a business strategy and plan}
Develop a strategy for sustainment as the research enterprise grows.
A business plan should describe sources of revenue such as licensing fees and
training courses. A strategy for recruiting additional collaborators should be
compatible with the governance system.  Others have suggested alternative models of 
funding from public sources, wherein software development, rather than research endeavors,
is funded for maintenance, refactoring and new features~\cite{de2014defining, downs2014recommendations}.

{\bf Be transparent}
Consider the structure of the organization. A productive research software community is very likely to be distributed across institutions and geography. Establishing GitHub repositories for documents, the
architecture, and test cases is just as important as posting the software.
Transparency and openness among the teams and organizations contributing and using 
the software platform is  necessary regardless of whether the software is
licensed as open source or not~\cite{Jansen20121495, darden}.

\section{Conclusion}\label{conclusion}
Over the last two years we have explored a cross section of scientific software
development organizations. We have identified a number of actions that
contribute to successful development. In this paper we have described some of
these actions and have given a very brief explanation as to why they are
effective at improving development and sustainment practices. Our study has
solidified our view that an explicit ecosystem strategy contributes to the
success of the software development project.

\section{Acknowledgments}\label{ack}
The work of the authors was funded by the National Science Foundation grant \#ACI-1343033.

\bibliographystyle{abbrv}
\bibliography{sigproc,science}
\end{document}